\documentstyle[11pt,epsf]{article}
\makeatletter
\newcommand{\artsectnumbering}{%
\@addtoreset{equation}{section}
\renewcommand{\theequation}{\thesection.\arabic{equation}}}
\makeatother  
\newcommand{\al}{\alpha}

\newcommand{\br}{\bar}
\newcommand{\bt}{\beta}
\newcommand{\bv}{\begin{verbatim}}
\newcommand{\Bx}{\Box}
\newcommand{\de}{\delta}
\newcommand{\ch}{\chi}
\newcommand{\ck}{\check}

\newcommand{\Dr}{\cal D}
\newcommand{\dt}{\dot}

\newcommand{\et}{\eta}
\newcommand{\ev}{\end{verbatim}}
\newcommand{\fr}{\frac}
\newcommand{\Ga}{\Gamma}
\newcommand{\ga}{\gamma}

\newcommand{\la}{\lambda}

\newcommand{\na}{\nabla}
\newcommand{\Om}{\Omega}
\newcommand{\om}{\omega}
\newcommand{\Ph}{\Phi}
\newcommand{\ph}{\varphi}
\newcommand{\ps}{\psi}

\newcommand{\Si}{\Sigma}
\newcommand{\si}{\sigma}
\newcommand{\sq}{\sqrt}
\newcommand{\ta}{\tau}

\newcommand{\Th}{\Theta}
\newcommand{\th}{\theta}
\newcommand{\Up}{\Upsilon}

\newcommand{\vb}{\verb}

\newcommand{\be}{\begin{equation}}
\newcommand{\ee}{\end{equation}} 
\newcommand{\eei}{\end{equation}\indent\indent}
\newcommand{\bc}{\begin{center}}
\newcommand{\ec}{\end{center}}
\newcommand{\ber}{\begin{eqnarray}}
\newcommand{\ear}{\end{eqnarray}}
\newcommand{\ba}{\begin{array}}
\newcommand{\ea}{\end{array}}

\newcommand{\p}{\partial}

\newcommand{\Hd}{{\cal H}}

\newcommand{\Lg}{{\cal L}}

\def\case#1/#2{\textstyle\frac{#1}{#2} }
\begin{document}
\title{Imploding Scalar Fields.}
\author{Mark D. Roberts, \\\\
Department of Mathematics and Applied Mathematics, \\ 
University of Cape Town,\\
Rondbosch 7701,\\
South Africa\\\\
roberts@gmunu.mth.uct.ac.za} 
\date{\today}
\maketitle
\vspace{0.1truein}
\bc Published:  {\it J.Math.Phys.} {\bf 37}(1996)4557-4573.\ec
\bc Eprint: gr-qc/9905006\ec
\bc Comments:  25 pages, no diagrams or tables,  LaTex2e.\ec
\bc 2 KEYWORDS:\ec
\bc Singularities:~~~Non-static spacetimes.\ec
\bc 1999 PACS Classification Scheme:\ec
\bc http://publish.aps.org/eprint/gateway/pacslist \ec
\bc 04.20Jb,  04.20Cv.\ec  
\bc 1991 Mathematics Subject Classification:\ec
\bc http://www.ams.org/msc \ec
\bc 83C15,  83C57.\ec
\newpage
\artsectnumbering
\begin{abstract}
Static spherically symmetric uncoupled scalar space-times have no
event horizon and a divergent Kretschmann singularity at the origin 
of the coordinates.   The singularity is always present so that 
non-static solutions have been sought to see if the singularities can 
develop from an initially singular free space-time.   In flat 
space-time the Klein-Gordon equation $\Box\ph=0$ has the non-static 
spherically symmetric solution $\ph=\si(v)/r$,  where $\si(v)$ is a once 
differentiable function of the null coordinate $v$.   In particular the 
function $\si(v)$ can be taken to be initially zero and then grow,  thus 
producing a singularity in the scalar field.   A similar situation 
occurs when the scalar field is coupled to gravity via Einstein's 
equations; the solution also develops a divergent Kretschmann 
invariant singularity,  but it has no overall energy.  To overcome 
this Bekenstein's theorems are applied to give two corresponding 
conformally coupled solutions.   One of these has positive ADM mass 
and has the properties:  i) it develops a Kretschmann invariant 
singularity,  ii)it has no event horizon,   iii)it has a well-defined
source,  iv)it has well-defined junction condition to Minkowski 
space-time, v)it is asymptotically flat with positive overall energy.
This paper presents this solution and several other non-static scalar 
solutions.   The properties of these solutions which are studied are 
limited to the following three:  i)whether the solution can be joined
to Minkowski space-time, ii)whether the solution is asymptotically 
flat,  iii)and if so what the solutions' Bondi and ADM masses are.
\end{abstract}
\section{Introduction.}
\label{sec:intro}
Singularities appear in many physical theories.   A singularity can be
defined as a domain where the description provided by the physical theory 
breaks down.   A prime example is the infinite electromagnetic potential of 
a point particle in Maxwell's theory.   A common approach to a theory which 
has singularities is to produce another theory governed by more general 
differential equations and then investigate whether the singularities still 
occur.   For example in electromagnetic theory Born and Infeld
investigated a generalization of Maxwell's Lagrangian to see if
the infinite potential was still present.   In gravitational theory the 
situation is more complex:  for a point particle the Kretschmann invariant
$R_{abcd}R_{....}^{abcd}$ diverges at the position of the particle,   
also on occasion the particle is surrounded by an event horizon.   
The event horizon is not singular in the sense defined above because a 
description of its effects can be made:  but the effects are so bizarre that 
along with singularities they can be called pathological.   
Vacuum general relativity frequently has both pathologies an example being 
the Schwarzschild solution for a point particle.   It is sometimes argued 
that the existence of both pathologies is palatable because the event horizon 
makes the divergent Kretschmann invariant invisible at infinity:  
but the physical description provided is still incomplete because it does not 
describe what happens at the divergent Kretschmann invariant.   
Vacuum general relativity does not furnish a good description of many 
astrophysical phenomena such as gravitational collapse because the pertinent 
space-time can contain many fields and fluids with non-vanishing stress.   
Astronomical observations purporting to be of "black holes" in fact do not 
directly observe event horizons,  the models which describe the situation 
merely use a steeper potential than that of Newtonian theory,  
the Newtonian limit of most relativistic theories produces such a potential.

To find exact solutions of gravitational field equations to fit a 
particular physical requirement is notoriously difficult.   An example is 
the two-body problem:  since the inception of general relativity the solution 
for two point particles acting only through gravity has been sought.   
Another example is the Yukawa problem:  in the absence of gravitation a 
massive scalar field has a shorter range than the corresponding massless case,
Yukawa's discovery of this lead to the postulation of nuclear forces.
How gravitation alters the shape of the potential is unknown,  and this
would be of experimental interest as the Yukawa potential can be accurately 
measured in accelerators;  also it is unknown whether the mass of the
interacting scalar field is the same as the ADM mass.   Scalar fields 
coupled to gravity produce unusual potentials,  complicated by the fact 
that the luminosity radial coordinate is often of the form $R=r~exp(\ph)$,
where the metric is explicitly expressed in terms of $r$.   Perhaps the 
simplest modification of vacuum general relativity is to choose a stress 
with an uncoupled scalar field.   When this is done the situation is 
mitigated;  for the static case there is no event horizon.   The problem
with static space-times is that the Kretschmann invariant is always 
present,  the space-time does not develop so as to produce it.   Non-static
spherically symmetric scalar solutions have been found,  one of which is
asymptotically flat,  
Roberts (1986)\cite{bi:mdr86}, (1989)\cite{bi:mdr89}.   
This solution has unusual energetics;  there is no overall energy,  
the positive energy of 
the scalar field and the negative energy of the gravitational field cancel 
out:  as the space-time develops energy is just exchanged between them.   
To overcome this Bekenstein's theorems are applied to uncoupled scalar field 
solutions to give two corresponding conformal scalar solutions.   One of 
these is asymptotically flat and has overall positive ADM energy.

In general relativity scalar fields can implode (and explode) like the 
example in the abstract;  the situation here is more pathological than 
in flat space because not only is it possible to produce a singularity 
in the scalar field but in addition there is a co-locational singularity 
of the gravitational field as indicated by the divergence of the Kretschmann
curvature invariant $R_{....}^{abcd}R_{abcd}$.   
Einstein-scalar space-times are sometimes not covered by theorems concerned 
with the general global and singular structure of space-time.  
These often assume the space-time stress tensor is restricted to: 
vacuum,  or electromagnetic stress,  or obey energy inequalities.   
An example of this is the formal definition of asymptotic flat space-times 
which assume that there is only an electromagnetic and gravitational 
field present.   Numerical studies show that the rate of 
decay of scalar fields is between these two Roberts(1986)\cite{bi:mdr86},
furthermore many spherical symmetric perfect fluid stresses do not have 
(in the sense of taking a radial coordinate $r\rightarrow\infty$)  
asymptotically flat space-times,  Roberts (1998)\cite{bi:mdr98}.   
Here a scalar solution is taken to be asymptotically flat if it reduces to 
Minkowski space-time as $r\rightarrow\infty$.   
The rate of decay of fields seems to be:
most fluids and some conformal scalars are not asymptotically flat,
then gravitation $>$ uncoupled scalars $>$ uncoupled vectors $>$ interacting 
fields.

In section \ref{sec:II} some Robertson-Walker scalar field 
solutions are given.   The examples chosen have scale factors 
which can be expressed in terms of straightforward functions.     
Ordinary scalar stresses obey the energy conditions but 
conformal scalar stresses sometimes do not.   Violation of the 
energy conditions allow the possibility of a singular freee 
Robertson-Walker space-time,  this can happen for conformal 
scalar fields Bekenstein and Meisels (1980)\cite{bi:BM},  and when 
cosmological constant is present,  Murphy (1973)\cite{bi:murphy}.   
Here the junction conditions of Robertson-Walker space-time are studied,  
and possible applications mentioned.   In section \ref{sec:III}Penny's 
(1976)\cite{bi:penny} solution is presented,  this can implode,  
but is not asymptotically flat.   In section \ref{sec:IV}the solution 
previously found by the author,  Roberts (1986)\cite{bi:mdr86} 
and (1989)\cite{bi:mdr89},  and also its two conformal scalar extensions 
as found by Bekenstein's theorems, are presented.   These solutions are 
explicitly dependent on a radial coordinate so that they are more similar 
to the example in the abstract than the examples in sections \ref{sec:II} 
and \ref{sec:III}.   
The solution and one of its Bekenstein extensions are asymptotically flat,  
and also have well defined junction conditions,  contrary to what has been 
stated by Szabodos (1990)\cite{bi:szabodos}.  The Bondi and ADM masses 
of the solutions are calculated.   The first two appendices derive 
the Clarke and Dray (1987)\cite{bi:CD} junction conditions subject 
to spherical symmetry.  In all the specific cases looked at here if a 
solution has a continuous metric (and hence first fundamental form) across 
a junction then its surface stress (which depends on the derivatives 
of the metric) vanishes and the junction is well defined.   The third 
appendix derives a the general expression for the ADM mass subject to 
spherical symmetry.

The field equations considered are Einstein's equations with scalar fields as 
source.   Specifically the ordinary scalar-Einstein equations are
\be
R_{ab}=2\ph_a\ph_b,
\label{eq:1.1}
\ee
where the coupling constant is taken to be contained in the scalar field $\ph$.
On occasion a null radiation field is also taken to be present with
\be
R_{ab}=\Ph^2k_ak_b,
\label{eq:1.2}
\ee
where $k_a$ is a null vector and $\Ph$ is a function of $x^a$.   
Conformal scalar solutions can be obtained from ordinary scalar solutions by 
using Bekenstein's (1973)\cite{bi:bekenstein} theorems.   
To derive these let barred quantities denote these quantities 
for an ordinary scalar solution, i.e. $\bar{R}_{ab}=2\ph_a\ph_b$.  
Then under a conformal transformation
\be
\bar{g}_{ab}=\Om g_{ab};
\label{eq:1.3}
\ee
the connection is transformed
\be
\bar{\Ga^a_{bc}}-\Ga^a_{bc}=\Om^{-1}(\de^a_b\Om_c+\de^a_c\Om_b-g_{bc}\Om^a);
\label{eq:1.4}
\ee
and the Ricci tensor is transformed
\be
\bar{R_{ab}}-R_{ab}=-2\Om(\Om^{-1})_{;ab}+\Om^{-2}(\Om^2)^c_{.c}g_{ab},
\label{eq:1.5}
\ee
where the covariant derivatives $";"$ are taken in the unbarred 
system.   Now take
\begin{equation}
\begin{array}{rcl}
\Om=\sqrt{\pm(1-2\xi^2\ph^2)}&=&\left\{
\begin{array}{l}
{\rm sech}\\
{\rm cosech}
\end{array}
\ \xi\ph\right.\\
\xi\ph=\sqrt{1\mp\Om^2}&=&\left\{
\begin{array}{l}
{\rm tanh}\\
{\rm coth}
\end{array}
\ \xi\ph\right.
\end{array}
\label{eq:1.6}
\end{equation}
where $\xi$ is a constant and $\ph$ is a function,   then substituting into 
the Ricci tensor \ref{eq:1.5},    obeys the equations for a conformal scalar 
field 
\be
(1/\xi^2-\ps^2)R_{ab}=4\ps_a\ps_b-2\ps\ps_{;ab}-(\ps\ps^c_.)_cg_{ab}.
\label{eq:1.7}
\ee
Thus given an ordinary scalar field solution \ref{eq:1.1} Bekenstein's 
theorems give two conformal scalar solutions \ref{eq:1.6};  Bekenstein 
refers to the upper sign conformal solution as type A and the lower sign 
conformal solution as type B,  as no confusion with blood types should occur 
the ordinary scalar solution is here called type O.   In the conventions used 
here,  the coupling constant is taken to be in the scalar field and thus
$\xi^2=\fr{1}{3}$.   The conformal scalar solutions are traceless and also 
obey $\ps^a_{.a}=0$. For type A the theorem generalizes for additional stress 
present,  in the case of a null radiation field this must transform as 
$\Ph^2\rightarrow\Om^{-2}\Ph^2$,  c.f.Bekenstein (1973)\cite{bi:bekenstein}
equation 15;   this generalization does not work for type B.   Type B solutions
are anticipated to have unusual global properties, for example as 
$\ph\rightarrow 0$:  $\Om^{-1}\rightarrow 0$,  $\ps\rightarrow\infty$;
and also as $\ph\rightarrow\infty$:  $\Om^{-1}\rightarrow\infty$,  
$\ps\rightarrow 1$;  but applying Bekenstein's theorems does not {\it a priori}
produce a maximally extended space-time so that exact solutions have to be 
specified before precise pronouncements on there global properties can be made.
Examples of static spherically symmetric conformal scalar fields with unusual 
properties can be found in Agnese and LaCamera (1985)\cite{bi:ALC}.

Having presented some of the properties of scalar fields we can now come back 
the questions:  general relativity and other gravitational theories are 
primarily macroscopic theories which couple to stresses that have macroscopic 
effect,  such electromagnetism and perfect fluids are natural choices -  why 
choose scalar fields?   First it only takes an infinitesimal scalar field to 
change the global nature of a space-time.   For example Wyman's solution,  
which is the general static spherically symmetric O-scalar-Einstien solution, 
is a two parameter solution $(M,\si)$ with $M$ the Schwarzschild mass and $\si$
the scalar charge,  an infinitesimal   is sufficient for there to be no event
horizon present.   Thus microscopic fields,  such as those of particle physics,
can have a global effect on space-time.   Secondly,  for $\ph$ time-like  an 
O-scalar field is a particular example of a perfect fluid.   Perfect fluids 
which are well-behaved and permeate the whole space-time can usually be shown 
to have no horizons,  Roberts (1998)\cite{bi:mdr98}.   The fluid 
conservation equations often allow the fluid index $\om$ to be equated 
with the fluid vector and hence the metric.   
Typically this results in equations such as the lapse 
$N=1/\om$;  thus a well-behaved fluid index   can often imply a well-behaved 
metric.   In appendix D an attempt is made to extend to fields the techniques 
that lead to this result.   Thirdly in microscopic physics hypothetical 
particles,  the Higgs scalars,  are used to introduce "mass" terms while 
preserving gauge invariance.   Although other mechanisms have been proposed,
for example by using fermion composites,  
or using fluids Roberts (1989)\cite{bi:mdr89},(1996)\cite{bi:mdr96},
the resulting Lagrangians have terms similar to scalar fields.   
Fourthly many quantum and unified theories have gravitational actions with 
terms of higher order.   The quadratic action can be split into two 
independent parts,  the traceless part being the Bach tensor and the other 
part the Pauli tensor.   Using a conformal factor solutions to these equations
can be found by Barrow and Cotsakis'(1988)\cite{bi:BC} method.   
The Bach tensor has several similarities to conformal scalar fields and there 
might be a theorem connecting their solutions.   Vacuum solutions can generate
O-scalar solutions by Buchdahl's trick (the analogous theorem for vector 
gauge theory is called the Julia-Zee correspondence) see for example 
Roberts (1986)\cite{bi:mdr86};  O-scalar solutions can generate A and B 
scalar solutions by Bekenstien's theorems;  and perhaps A and B scalar 
solutions can generate Bach-Einstien solutions.   Fifthly O-scalar solutions 
obey the energy conditions.   The energy conditions are inequalities designed 
to judge whether macroscopic fluids have reasonable energetics:  they break
down when considering the interacting fields necessary for particle physics,  
see for example Hawking and Ellis (1973)p.95 \cite{bi:HE}.

Apart from the above five reasons for investigating scalar fields 
they can be viewed as merely a scalar function defined on a region of space,  
and such a requirement seems to be fairly ubiquitous in physics.   Systematic 
discussion of them is hindered because there is no recognized way of 
classifying them. Some indications of their properties are given by their:- 
{\it Coupling classification},  call scalar fields coupled only to 
Einstein's equations type O:  conformal scalar fields coupled to Einstein's 
equations type C,  scalar fields with mass self-interaction type Y,  
renormalisable scalar fields with fourth order self-interaction type l,   
inflationary scalar fields with potential $V(\ph)$ type V,   
scalar fields that can be represented as fluids type F,   
symmetry breaking scalar fields coupled to vector fields type H - and so on.
{\it Generational classification}:   exact 
scalar field solutions can be generated from exact solutions to simpler 
differential equations.   For a given configuration usually the generated 
solution is not the most general one.   
{\it Stress classification}:  stress tensors
can be classified by the Segre or Plebanski methods.   
{\it Energy classification}:
for a space-time with Lorenz signature (-,+,+,+) rather than positive
definite signature (+,+,+,+),  vectors can be time-like,  null, or space-like.
The existence of time-like vectors allows measures of the energy to be defined.
For a given space-time $\ps_a$ and $\ps_bT^b_{.a}$ are vectors that can be 
time-like,  null,  or space-like;  the energy conditions can be investigated 
and on occasion the overall energy measured.

For conformal scalar fields the energy conditions can be complicated;  
hand calculations of them for the specific solutions presented here are too 
long to be practicable.   Three observations are now stated which give a rough
indication of what energetics to expect.   The first observation is that the 
general stress for conformal scalar fields contains terms of undetermined 
sign and this remains the case even if the conformal scalar stress has been 
obtained by using Bekenstein's theorems;  this can been seen from \ref{eq:1.6}
and \ref{eq:1.7}.   The second observation is for an O-scalar solution it is 
possible to consider whether the vector $\ph_a$ is space-like,  time-like, 
or null:  this just depends on the Ricci scalar because  
$\bar{g}^{ab}_{..}=\fr{1}{2}\bar{R}$;  now using Bekensteins' theorems both 
for type A and B there is the equation  
$g^{ab}_{..}\ps_a\ps_b=\fr{1}{2}\bar{R}$.  
Thus there is no change in the causal direction of the scalar field.    
The third observation is achieved by direct inspection of the scalar fields.
Neglect that the equation for the stress of a conformal scalar field differs 
from that of the ordinary case and also that Bekenstein's theorems introduce 
a conformal factor into the metric,  then the energy conditions will just 
depend on the relative size of the scalar fields involved.  
The type O scalar field is a negative real scalar quantity,  
Bekenstein's theorems give that the type A 
scalar is the tanh of this and that the type B is the coth of this,  thus the 
scalar fields take real values such that $0>A>O>B>-\infty$.   Now the type O 
scalar field on occasion (an example being that of section IV)  contains the 
same quantity of energy,  but of the opposite sign as the gravitational field;
the above inequalities suggest that the type A solution would have total 
positive energy because the scalar field is not so negative,  and also that 
the type B solution would have total negative energy.   This is found to be 
the case for the type A solution described in the conclusion.  The above 
suggests that it is a reasonable guess that type A solutions have well defined
energy conditions and that type B do not:  this is what would be expected from
the known exact solutions where it would account for type A have mundane 
properties whereas those of type B solutions are bizarre.
\section{Robertson-Walker Scalar Solutions.}
\label{sec:II}
The Robertson-Walker line element can be put in the form
\be
ds^2=-N(t)^2dt^2+R(t)^2d\Si^2_{3,k},
\label{eq:2.1}
\ee
where
\ber
d\Si^2_{3,k}=d\ch^2+f(\ch)^2(d\th^2+sin^2(\th)d\ph^2),\nonumber
\ear
and
\ber
f(\ch)=sin(\ch),\ch,sinh(\ch),\nonumber
\ear
for $k=+1,0,-1$ respectively.   Taking $A=R^2,B=R^2f^2,C=N^2$,
and $\ch=r$ gives the line element in the spherically symmetric form 
\ref{eq:A2.1}.   $N$ is called the lapse and R the scale factor.   
$N$ can be absorbed into the line element,  the choice $N=1$ gives the 
Robertson-Walker line element in proper time.  For the choice $N=R$ 
Robertson-Walker space-time is conformal to the Einstein static universe 
and by convention the time coordinate is denoted by $\et$.  For $N=1$
the scale factor $R$ can be expanded as a Taylor series around 
a fixed time $t=t_0$ thus
\be
R=R_0[1+H_0(t-t_0)-\fr{1}{2}q_0H_0^2(t-t_0)^2+O(t-t_0)^3],
\label{eq:2.2}
\ee
where 
\be
H\equiv\dot{R}/R,
\label{eq:2.3}
\ee
is called the Hubble parameter and
\be
q\equiv\ddot{R}.R/(\dot{R}^2),
\label{eq:2.4}
\ee
is called the deceleration parameter,  the subscript $"0"$ indicates that the 
parameter is evaluated at $t=t_0$,  and $\dot{R}=\p R$.   

The equation of state
\be
p=(\ga-1),
\label{eq:2.5}
\ee
produces equations equivalent to those of an ordinary scalar field in the 
particular case of $\ga=2$ (see appendix D).   Einstein's equations have been 
solved by Vajk(1968)\cite{bi:vajk} for the metric \ref{eq:2.1} and 
equation of state \ref{eq:2.5},  specializing to the $\ga=2$ ordinary scalar 
field case gives  
\ber
&&k=0,~~~\Xi=\al\et^{1/2},~~~\ph=\fr{\sqrt{3}}{2}ln~\et,\nonumber\\
&&k=+1,~~~\Xi=\al(sin~\et~ cos~\et)^{1/2},~~~
\ph=\fr{\sqrt{3}}{2}ln~tan~\et,\\
&&k=-1,~~~\Xi=\al(sinh~\et~ cosh~\et)^{1/2},~~~
\ph=\fr{\sqrt{3}}{2}ln~tanh~\et,\nonumber
\label{eq:2.6}
\ear
where
\ber
\al=2R_0\sqrt{\fr{H_0^2R_0^2}{c^2}+k},\nonumber
\ear
$c$ is the speed of light and $\Xi$ is equal to both the scale factor 
and the lapse,  i.e. $\Xi=N=R$.   The $k=0$ solution is one of the 
few solutions known to have an exact form for the world function,  
Roberts (1993)\cite{bi:mdr93}.

Applying Bekenstein's theorems
\ber
&&k=0,~~~\Up=\fr{\al}{2}(\et\pm1),~~~
\fr{1}{\sqrt{3}}\ps=\fr{\et\mp1}{\et\pm1},\nonumber\\
&&k=+1,~~~\Up=\fr{\al}{2}(sin~\et\pm cos\et),~~~
\fr{1}{\sqrt{3}}\ps=\fr{sin~\et\mp cos\et}{sin~\et\pm cos~\et},\\
&&k=-1,~~~\Up=\fr{\al}{2}exp(\pm\et)~~~,
\fr{1}{\sqrt{3}}\ps=-exp(\mp\et),\nonumber
\label{eq:2.7}
\ear
where $\Up=\Xi\Om^{-1}$.   In the $k=0$ case the $\pm1$ can be absorbed into
the line element by defining $\et'=\et\pm1$,  giving a single solution.   
Transferring the $k=+1$ solution to proper time by defining 
$t=(\al/2)(-cos~et\pm sin~\et)$ gives 
\be
ds^2=-dt^2+\left(\fr{\al^2}{2}-t^2\right)d\Si^2_{3,+1},~~~
\fr{1}{\sqrt{3}}\ps=\left(\fr{\al^2}{2t^2}-1\right),
\label{eq:2.8}
\ee
showing that there is only one $k=+1$ metric.   In the $k=-1$ case define 
$t=(\al/2)exp(\pm\et)$ to give
\be
ds^2=-dt^2+t^2d\Si^2_{3,-1},~~~
\fr{1}{\sqrt{3}}\ps=-\fr{\al^2}{4t^2}.
\label{eq:2.9}
\ee
this is just the Milne universe,  further discussed at equation \ref{eq:2.16};
the field $\ps$ is a ghost field that does not contribute to the stress.   
Conformal scalar stresses are traceless,  this can be used to reduce the 
number of equations,  in particular the Einstein-conformal scalar equations 
with Robertson-Walker metric can be quickly reduced to one equation 
\be
R_{,\et\et}=-kR,
\label{eq:2.10}
\ee
this gives solutions more general than those of \ref{eq:2.7},  however they 
are particular instances of the conformal scalar and incoherent radiation 
solutions of Bekenstein (1974)\cite{bi:bekenstein}\vb+#+6.  
Bekenstein's theorems can then be used in reverse to give generalizations 
of \ref{eq:2.6}.

The null junction conditions are studied by defining
\be
v=\et+r.
\label{eq:2.11}
\ee
Robertson-Walker space-time in the conformal time coordinate $\et$ takes the 
single null form \ref{eq:A1.1} with $X=S=Yf^{-2}=R(v-r)^2$.   
The $\th$ and $\ph$ components of the surface stress vanish identically,  
the $v$ and $r$ components are given by
\be
\ta_{ab}=-f^{-2}R^{-2}[(R^2f^2)']n_an_b,
\label{eq:2.12}
\ee
which do not vanish at a junction with Minkowski space-time.    

The non-null junctions
\footnote{footnote added 1999:  junctions between Schwarzschild 
space-tme and the pressure-free Friedman universe are discussed in 
Stephani,H.General Relativity,  An introduction to the theory of the 
gravitational field,  Cambridge University Press (1982)\S 27.3}
are studied by calculating the second fundamental form 
as in Appendix B.   The second fundamental form across the space-like surface 
normal to \ref{eq:A2.3} is
\ber
&&K_{rr}=-\fr{R\dot{R}}{N},\label{eq:2.13}\\
&&K_{\th\th}=sin^{-2}(\th)K_{\ph\ph}=-\fr{f^2R\dot{R}}{N},\nonumber
\ear
which gives no junctions to Minkowski space-time.   The surface normal
to the radial space-like vector \ref{eq:A2.5} has second fundamental form
\ber
&&K_{tt}=0,\label{eq:2.14}\\
&&K_{\th\th}=sin^{-2}(\th)K_{\ph\ph}= ff'R.\nonumber
\ear
Again there are no junctions to Minkowski space-time.  The radial space-like 
vector is not well suited to Robertson-Walker geometries,  choosing the 
isotropic space-like vector \ref{eq:A2.7} gives second fundamental form
\ber
&&K_{tt}=K_{tr}=0,\nonumber\\
&&K_{r\th}=\fr{K_{t\ph}}{sin~\th}
          =R\fr{K_{\th\ph}}{sin~\th}
          =-\fr{1}{2}fK_{rr}
          =\fr{-f'R}{3\sqrt{3}},\\
&&K_{\th\th}=\fr{K_{r\th}}{sin~\th}
            =\fr{f}{3\sqrt{3}}\{-cot~\th+2f'R\}.\nonumber
\ear
Again there are no junctions to Minkowski space-time. 

Consider Minkowski space-time in the form \ref{eq:2.1} with $k=0$ and 
$N=R=1,f=r$,  and apply the coordinate transformation
\be
t=\bar{t}cosh~\ch,~~~  
r=\bar{t}sinh~\ch.
\label{eq:2.16}
\ee
This transformation gives the Milne universe which has $N=1, R=t$,  and 
$f=sinh~ r$.   The Milne universe is flat and is identical to Minkowski 
space-time except that there is a point removed at the origin $t=0$.   
At first sight it might be expected that the Milne universe could be 
joined to Minkowski space-time across the surface chosen here.   
The reason that this does not happen is that the space and time coordinates 
have been "mixed" by \ref{eq:2.16} so that if there was a well defined 
junction it would be across a different surface from those chosen here.   
A general treatment of redefinitions of space and time coordinates in 
Robertson-Walker space-time can be found in 
Infeld and Schild (1945)\cite{bi:IS}.

The junction conditions of Robertson-Walker space-time have two further 
applications.   The first is the production of a spherical Minkowski cavity 
which has implications for Mach's principle,  
see for example Weinberg (1972)\cite{bi:weinberg}p.474.  
A point inside the cavity is an inertial frame if it does 
not rotate with respect to the reference frame of the rest of the Universe,  
which is taken to be given by the Robertson-Walker space-time surrounding the 
cavity.   A different approach to Mach's principle is discussed in Roberts 
(1985)\cite{bi:mdr85}.   The second is to the cell universe models.    
The surface normal to the vectors chosen here do not allow junctions between 
Robertson-Walker space-time and Schwarzschild space-time,  
thus for the Schwarzschild cell universe of 
Lindquist and Wheeler (1957)\cite{bi:LW} to work a different vector has 
to be chosen or different physical assumptions made.
\section{Penny's Solution}
\label{sec:III}
Penny's solution (1976)\cite{bi:penny},  and the related solutions of 
Gurses (1977)\cite{bi:gurses} and Ray (1977)\cite{bi:ray} are 
conformally flat.   The conformal factor generating technique used to 
find these solutions is also used to study solutions of higher order 
gravity theories,  Barrow and Cotsakis (1988)\cite{bi:BC}.   Here attention 
is restricted to Penny's solution where the conformal factor and the ordinary 
scalar field are given by
\be
\Xi=k_ax_.^a+a,~~~          
k_{a,b}=0,~~~
\ph=\sqrt{3}ln\Xi,
\label{eq:3.1}
\ee
respectively.   Defining 
\be
K_a=-\Xi k_a,
\label{eq:3.2}
\ee
$K_a$ is a Killing vector which is null iff $k_a$ is null.   
The conformal factor can be expressed as
\be
\Xi^2=a+bt+cx+dy+ez,
\label{eq:3.3}
\ee
where $a,b,c,d$ are constants.   There is no asymptotically flat solution.   
For $a=c=d=e=0,~~b=2R^2_0H_0$,   this is the $k=0$ solution \ref{eq:2.6}.

Using Bekenstein's theorems ,  conformal scalar solutions are
\ber
&&\Up=\Om^{-1}\Xi,\nonumber\\
&&2\Om=2\left\{
\ba{rl} &cosh\\&sinh\ea\right.
\xi\ph=\Xi\pm\Xi^{-1},\\
&&\xi\ps=\left\{
\ba{rl} &tanh\\&coth\ea\right.
\xi\ph=\fr{\Xi^2\mp1}{\Xi^2\pm1},\nonumber
\label{eq:3.4}
\ear
giving
\ber
\Up=2\Xi^2/(\Xi^2\pm1).\nonumber
\ear
Defining 
\be
K_a=\Up^2k_a,
\label{eq:3.5}
\ee
again $K_a$ is a Killing vector which is null iff $k_a$ is null.
\section{The solution previously found by the Author}
\label{sec:IV}
The solution found in Roberts (1986)\cite{bi:mdr86} 
and (1989)\cite{bi:mdr89} is
\be
ds^2=-(1+2\al_{,v})dv^2+2dvdr+r(r-2\al)d\Si^2_2,
\label{eq:4.1}
\ee
where $d\Si^2_2=d\th^2+sin^2\th~d\ph$,  and $\al$ is a twice differentiable 
function of $v$.   The stress is given by a scalar field and a null radiation 
field
\be
\ph=\fr{1}{2}ln\left(1-2\fr{2\al}{r}\right),~~~
\Ph^2=\fr{2\al\al_{,vv}}{r(r-2\al)}.
\label{eq:4.2}
\ee
Defining the luminosity distance $R^2=r(r-2\al)$,  and taking the positive 
sign of the square root the solution becomes
\ber
&&ds^2=\left(-1+\fr{2\al\al_{,v}}{\la}\right)dv^2
       +\fr{2R}{\la}dRdv+R^2d\Si^2_2,\label{eq:4.3}\\
&&\ph=\fr{1}{2}ln\left(\fr{\la-\al}{\la+\al}\right),~~~
\Ph^2=2\fr{\al\al_{,v}}{R^2},\nonumber
\ear
where $\la^2=\al^2+R^2$.   The Bondi mass $M(v)$ is half the coefficient 
of the $R^{-1}$ term of $g_{vv}$,  expanding gives
\be
M(v)=\al\al_{,v}.
\label{eq:4.4}
\ee
For the null radiation field to vanish $\al_{,vv}=0$ or $\al=\si v+\bt$,
where $\si$ and $\bt$ are constants.   It is straightforward to show that $\bt$
can be absorbed into the line element leaving $\al=\si v$;  this can be 
substituted into \ref{eq:4.3} for a form of the metric using the luminosity 
radial coordinate,  alternatively it can be substituted into \ref{eq:4.1} 
giving
\ber
&&ds^2=-(1+2\si)dv^2+2dvdr+r(r-2\si v)d\Si^2_2,\label{eq:4.5}\\
&&\ph=\fr{1}{2}ln\left(1-\fr{2\si v}{r}\right).\nonumber
\ear
Defining $v'=\sqrt{1+2\si v}$,  and $r'=r/\sqrt{1+2\si}$ gives
\ber
&&ds^2=-dv'^2+2dv'dr'+r'((1+2\si)r'-2\si v')d\Si^2_2,\label{eq:4.6}\\
&&\ph=\fr{1}{2}ln\left(1-\fr{2\si}{(1+2\si)}\fr{v'}{r'}\right)\nonumber
\ear
then defining $t'=v'-r'$ gives
\ber
&&ds^2=-dt'^2+dr'^2+r'(r'-2\si t')d\Si^2_2,\nonumber\label{eq:4.7}\\
&&\ph=ln(R/r).
\ear
also defining $t=t'/1+2\si$ the solution can be put in the form
\ber
&&ds^2=-(1+2\si)dt^2+\fr{dr^2}{(1+2\si)}
       +r\left(\fr{r}{(1+2\si)}-2\si t\right)d\Si^2_2,\nonumber\\
&&\ph=\fr{1}{2}ln\left(\fr{1}{(1+2\si)}-\fr{2\si t}{r}\right).
\label{eq:4.8}
\ear
and from this form using \ref{eq:A3.5} the ADM mass is seen to vanish 
identically.

Using Bekenstein's theorems to find conformal scalar solutions \ref{eq:4.3} 
gives
\ber
&&R=\bar{R}\Om^{-1}
   =\bar{R}\left\{
\ba{lr}&cosh\\&sinh\ea\right.\xi\ph
   =\fr{1}{2}\bar{R}^{1-\xi}[(\la-\al)^\xi\pm(\la+\al)^\xi],\nonumber\\
&&\xi\ph=\fr{(\la-\al)^\xi\mp(\la+\al)^\xi}
            {(\la-\al)^\xi\pm(\la+\al)^\xi},
\label{eq:4.9}
\ear
where $\la^2=\bar{R}^2+\al^2$ and $\bar{R}$ denotes the luminosity coordinate 
for the O-scalar solution,  and $R$ denotes it for the conformal scalar 
solution.   For the type A solution general can be retained if the null 
radiation field is transformed, but for the type B solution $\al$ 
must equal $\si v$.  Inspecting \ref{eq:4.9} gives limiting behaviour 
of the conformal scalar solution in terms of the luminosity coordinate 
for the ordinary solution,
\ber
\bar{R}\uparrow\infty,~~{\rm Type A:}~~R\uparrow\infty,~~\ps\uparrow0&,&
        {\rm Type B:}~~R\downarrow-\xi,~~\ps\downarrow-\infty,\nonumber\\
\bar{R}\downarrow0,~~{\rm Type A:}~~R\downarrow0,~~\ps\downarrow-1&,&  
                {\rm Type B:}~~R\downarrow-0,~~\ps\uparrow-1.
\label{eq:4.10}
\ear
The type B solution does not have an asymptotically flat region,  
so that attention is restricted to the type A solution.   
Expanding \ref{eq:4.9} for large $R$ gives
\be
R=\bar{R}\left(1+\fr{\xi^2\al^2}{2\bar{R}^2}
               +\large O\left(\fr{\al}{\bar{R}}\right)^3\right).
\label{eq:4.11}
\ee
Solving this quadratic
\be
\bar{R}=\fr{R}{2}\left(1\pm\sqrt{1-\fr{2\xi^2\al^2}{R^2}}\right),
\label{eq:4.12}
\ee
the upper sign is taken so that $R=\bar{R}$ when $\xi=0$.   Differentiating
\be
d\bar{R}=\fr{1}{2}\left(1+\fr{2}{\sqrt{1-2\xi^2\al^2/R^2}}\right)
         -\fr{2\xi^2\al\al_{,v}dv}{R\sqrt{1-2\xi^2\al^2/R^2}},
\label{eq:4.13}
\ee
inserting into the line element and using \ref{eq:4.3},\ref{eq:4.12},
\ref{eq:4.13} and that $\la^2=\al^2+\bar{R}^2$ gives 
\be
g_{vv}=-1+\fr{2(1-\xi^2)}{R}\al\al_{,v}+\large O(R^{-2}).
\label{eq:4.14}
\ee
Thus the Bondi mass is given by
\be
M(v)=(1-\xi^2)\al\al_{,v},
\label{eq:4.15}
\ee
which is two thirds of \ref{eq:4.4}.  To calculate the ADM mass note that the 
conformal factor can be used
\be
A_.^t=\bar{A}_.^t+\lim_{r\rightarrow\infty}
      \left[-\fr{1\bar{B}}{2r}(\Om^{-2})'\right],
\label{eq:4.16}
\ee
where $\bar{B}$,  $\bar{A}$ are the values of these quantities in the O-scalar 
solution,  using the metric in the form \ref{eq:4.8} and noting that $\bar{A}$
vanishes
\be
A_.^t=\lim_{r\rightarrow\infty}-\fr{\bar{B}\xi}{2}sinh(2\xi\ph)\cdot\ph'    
     =\lim_{r\rightarrow\infty}-\fr{\xi\ph}{2}sinh(2\xi\ph)\cdot t.
\label{eq:4.17}
\ee
expanding $sinh(2\xi\ph)$ for $r\rightarrow\infty$ gives
\be
A_.^t=\left\{
\ba{rl}&\xi^2\si^2t,~~~~~~~{\rm for}-1<2\si<1,\\
       &2\xi^{xi-2}\si^{-\xi+1}t,~~{\rm for} 2\si\geq1,
\ea\right.
\label{eq:4.18}
\ee
for $2\si\geq-1$ the signature of the metric is not retained.

The surface stress \ref{eq:A1.6} must vanish at any null junction;  
this implies $[Y']$ must vanish if $\ta_{vv},  \ta_{vr},  \ta_{rv}$,  
and $\ta_{rr}$ are to vanish,  and $[(X/S)'/2S]$ must vanish if $\ta_{\th\th}$,
and  $\ta_{ph\ph}$ are to vanish.   For the type O solution take the line 
element \ref{eq:4.3},  $[Y']$ vanishes as this line element is expressed in 
terms of the luminosity radial coordinate already and
\be
\fr{1}{2S}\left(\fr{X}{S}\right)'=\fr{\al}{2R^{\xi}}(\al-2\la\al_{,v}).
\label{eq:4.19}
\ee
Now the metric can be chosen to continuously join to Minkowski space-time by 
taking $\al$ to continuously increase from (or decrease to) $\al=0$;  
\ref{eq:4.19} shows that there is no surface stress at the join $\al=0$ so that
the field equations are obeyed throughout the space-time.   For the type A and
B solutions \ref{eq:4.9} gives
\ber
&&Y=\fr{1}{4}\bar{R}^{2-2\xi}\{(\la-\al)^\xi\pm(\la+\al)^\xi\},\label{4.20}\\
&&Y'=\fr{1}{2}\bar{R}^{1-2\xi}\{(\la-\al)^\xi\pm(\la+\al)^\xi\}\nonumber\\
&&   \times\left\{(1-\xi)((\la-\al)^\xi\pm(\la+\al)^\xi)
                +\fr{\xi}{\al}\bar{R}^2((\la-\al)^{\xi-1}
                     \pm(\la+\al)^{\xi-1})\right\},\nonumber\\
&&\fr{1}{2S}\left(\fr{X}{S}\right)'=\fr{4\al(\al-\la\al_{,v})}
              {\bar{R}^{3-2\xi}\{(\la-\al)^\xi\pm(\la+\al)^\xi\} }.\nonumber
\ear
For the type B solution the metric is not continuous at $\al=0$ as would be 
anticipated from the general remarks in the introduction.   The type A solution
again has a metric which can be chosen to continuously join to Minkowski 
space-time by taking $\al$ to behave as before.

At any junction across a time-like surface the limits \ref{eq:A1.7} of the 
second fundamental form either side of the junction must coincide.   For the 
type O solution,  dropping the primes on the metric \ref{eq:4.7} gives the 
second fundamental form \ref{eq:A2.6}
\be
K_{\th\th}=sin^{-2}\th K_{\ph\ph}=r-\si t,
\label{eq:4.21}
\ee
which gives a junction where the field equations are defined with Minkowski 
space-time at t=0.  For type A and B conformal solutions the extension of the 
metric \ref{eq:4.7} has second fundamental form \ref{eq:A2.6}
\ber
&&K_{tt}=-\xi\si t(r(r-2\si t))^{-\xi/2-1}\cdot\{(r-2\si t)^\xi\mp r^\xi\},
                                                              \label{eq:4.22}\\
&&K_{\th\th}=sin^{-2}K_{\ph\ph}=\fr{1}{2}(r(r-2\si t))^{-\xi/2}\nonumber\\
&&~~~~~~~~~~~\cdot\{(r+(\xi-1)\si t)(r-2\si t)^\xi\pm(r-(\xi+1)\si t)r^\xi\}.
                                                               \nonumber
\ear
Again there is a junction at $t=0$.
\section{Conclusion}
\label{sec:conc}
Solutions to the Einstein-scalar equations which can represent an imploding 
scalar field were presented.   Bekenstein's theorems were used to generate 
the corresponding Einstein-conformal scalar solutions.   The Robertson-Walker 
solutions presented here are examples of solutions previously found by Vajk 
and Bekenstein;  they are not asymptotically flat and cannot be joined to 
Minkowski space-time by the methods used here.   Penny's solution also can 
represent an imploding scalar field but it is not asymptotically flat,  
and only when it reduces to a Robertson-Walker metric is it spherically 
symmetric.  The solution previously found by the author,  and its' Bekenstein 
type A extension,  are asymptotically flat and have well defined junctions 
with Minkowski space-time,  and therefore can represent a scalar field 
imploding from nothing,   thus generalizing the example in the abstract.   
This solution has Bondi mass $\al\al_{,v}$,   and zero ADM mass,  the zero ADM
mass is because the energy of the gravitational field is negative and equals 
the positive energy of the scalar field.   The type A extension has Bondi mass
$(1-\xi^2)\al\al_{,v}$,  and ADM mass $\xi^2\si^2 t,(|2\si|<1)$.   Assuming 
that the null radiation field vanishes,  so that there is only the conformal 
scalar field present the Type A solution has $\al=\si v$,  therefore
\newpage
\ber
&&{\rm Type O:}~~~M(v)=\si^2 v,~~~A_.^t=0,\nonumber\\
&&{\rm Type A:}~~~M(v)=\fr{2}{3}\si^2 v,~~~A_.^t=\fr{1}{3}\si^2 t.\nonumber
\ear
The type A solution might violate the energy conditions,  but subject to this 
proviso it is possible to start with Minkowski space-time and join the type A 
solution at $t=0$ generating non-zero ADM mass.   This only goes to show that 
you can get something (as measured by ADM mass) from nothing.
\appendix
\section{Appendix A: Junction Conditions Across A Null Surface}
\label{sec:A}
In single null coordinates a spherically symmetric line element can be 
written as
\be
ds^2=-X~ dv^2+2S~ dvdr+ Y(d\th^2+sin^2\th d\ph^2).
\label{eq:A1.1}
\ee
A suitable null tetrad is
\be
l_a=(S,0,0,0),~~~  
n_a= (X/2S,-1,0,0),~~~
m_a=(0,0,1,i sin\ph)\sqrt{Y/2}.
\label{eq:A1.2}
\ee
The projection tensor is defined by $q_{ab}=g_{ab}+2l_{(a}n_{b)}$,  and has
non-vanishing components $q_{\th.}^{~\th}=q_{\ph.}^{~\ph}=1$.   The internal 
second fundamental form $\ch_{ab}=n_{d;c}q_{.a}^dq_{.b}^c$, involves covariant
derivatives of $n_a$ which can be calculated using the Christoffel symbols in 
Roberts(1989)\cite{bi:mdr89},  and it has non-vanishing components
\be
\ch_{\th\th}=sin^{-2}\th\ch_{\ph\ph}=-(XY'/2S+Y_{,v})/2S.
\label{eq:A1.3}
\ee
the external second fundamental form $\ps_{ab}=l_{d;c}q_{.a}^cq_{.b}^d$ is
\be
\ps_{\th\th}=sin^{-2}\th\ps_{\ph\ph}=Y'/2,
\label{eq:A1.4}
\ee
and the normal fundamental form $\et_a=l_{d;c}q_{.a}^c n_.^d$ vanishes.   
The surface gravity $\om=-l_d n_.^c n_{.;c}^d$ is
\be
\om=-(X/S)'/2S.
\label{eq:A1.5}
\ee
The surface stress $\ta_{ab}=-[Tr\ps]n_an_b-2[\et_{(a}]_{b)}-[\om]q_{ab}$ is
\be
\ta_{ab}=-[Y']n_an_b/Y+[(X/S)'/2S]q_{ab},
\label{eq:A1.6}
\ee
with $n_a$ and $q_{ab}$ given by \ref{eq:A1.2} and where the bracket "[  ]" is
defined by
\be
[Q]|_y=\lim_{x\rightarrow y^+}Q-\lim_{x\rightarrow y^-}Q,
\label{eq:A1.7}
\ee
for a point $y$ in the surface.
\section{Appendix B:  
         Junction Conditions Across Space-like And Time-like Surfaces}
\label{sec:B}
The line element can be taken in the form
\be
ds^2=-C~ dt^2+A~ dr^2+B(d\th+sin^2\th d\ph^2).
\label{eq:A2.1}
\ee
For a non-null surface,  the surface stress vanishes iff the second 
fundamental form obeys
\be
[K_{ab}]=0,
\label{eq:A2.2}
\ee
where the bracket "[ ] " is defined by \ref{eq:A1.7}.   A suitable unit 
time-like vector field
\be
U_a=(\sqrt{C},0,0,0).
\label{eq:A2.3}
\ee
The projection tensor $h_{a.}^b=g_{a.}^b+U_aU_.^b$,  has three components 
$h_{.r}^r=h_{\th .}^\th=h_{\ph .}^\ph=1$  The second fundamental form is 
$K_{ab}=U_{c;d}h_{.a}^ch_{.b}^d$ it involves covariant derivatives of $U_a$   
which can be calculated using the Christoffel symbols in 
Roberts (1989)[2]\cite{bi:mdr89},  the non-vanishing components are
\be
K_{rr}=-\dot{A}/(2\sqrt{C}),~~~ 
K_{\th\th}=sin^{-2}\th K_{\ph\ph}=-\dot{B}/(2\sqrt{C}),
\label{eq:A2.4}
\ee
where $\dot{A}=\p_tA$.   The radial unit space-like vector is
\be
U_a=(0,\sqrt{A},0,0),
\label{eq:A2.5}
\ee
The space-like projection tensor $h_{a.}^{~b}=g_{a.}^{~b}-U_aU_.^b$,  
has three components $h_{t.}^{~t}=h_{\th.}^{~\th}=h_{\ph.}^{~\ph}=1$.   
The second fundamental form for the corresponding time-like surface is
\be
K_{tt}=-C'/(2\sqrt{A}),~~~
K_{\th\th}=sin^{-2}\th K_{\ph\ph}=B'/(2\sqrt{A}),
\label{eq:A2.6}
\ee
where $C'=\p-r C$.   Choosing the isotropic unit space-like vector
\be
U_a=\fr{1}{\sqrt{3}}(0,\sqrt{A},\sqrt{B},\sqrt{B}~sin~\th),
\label{eq:A2.7}
\ee
similarly the non-vanishing components of the second fundamental form are
\ber
&&K_{tt}=-C'/(2\sqrt{3}A),\label{A2.8}\\
&&K_{tr}=-2\sqrt{\fr{A}{B}}K_{t\th}
        =\fr{-2}{sin~\th}\fr{A}{B}K_{t\ph}
        =\fr{-\dot{A}}{3\sqrt{3A}}+\fr{\dot{B}}{3B}\sqrt{A}{3},\nonumber\\
&&K_{r\th}=\fr{K_{r\ph}}{sin~\th}
          =\fr{\sqrt{A}K_{\th\ph}}{sin~\th}
          =\fr{-1}{2}\sqrt{\fr{B}{A}}K_{rr}=\fr{-B'}{6\sqrt{3B}},\nonumber\\
&&K_{\th\th}=\fr{K_{r\th}}{sin~\th}
            =\fr{1}{3\sqrt{3A}}\{-B~ cot~\th+B'\}.\nonumber
\ear
The above three vectors \ref{eq:A2.3}, \ref{eq:A2.5},  and \ref{eq:A2.7} are 
suitable for the majority of purposes;  but for example,  if there is "Mixing"
between the space and time coordinates,  like that of equation \ref{eq:2.16}
then other vectors have to be used.
\section{Appendix C: The ADM Energy}
\label{sec:C}
The ADM energy for a spherically symmetric space-time is found by generalizing
Weinberg's(1972)\cite{bi:weinberg} derivation for the Schwarzschild 
solution.   Define the rectilinear coordinates
\be
x_1=r~ sin~\th cos~\ph,~~~  
x_2=r~ sin~\th sin~\ph,~~~
x_3=r~ cos~\th,
\label{eq:A3.1}
\ee
the line element \ref{eq:A2.1} becomes
\be
ds^2=-C~dt^2+(A/r^2-B/r^4)(\underline{dx.x})^2+(B/r^2)\underline{dx}^2.
\label{eq:A3.2}
\ee
Defining $h_{ij}=g_{ij}-\et_{ij}$,  and $n-.^i=x_.^i/r,  i,j= 1,2,3$, gives
\be
h_{ij}=(A-B/r^2)n_in_j+(B/r^2-1)\de_{ij}.
\label{eq:A3.3}
\ee
The ADM mass is given by 
\be
A_.^t=\oint dS_.^i(h^j_{.i,j}-h^j_{.j,i}).
\label{eq:A3.4}
\ee
using \ref{eq:A3.3} this is
\be
A_.^t=\fr{-1}{2}r\left(\fr{-A}{r}+\left(\fr{B}{r^2}\right)'+\fr{B}{r^3}\right).
\label{eq:A3.5}
\ee
The remaining components of the ADM vector are given by 
\be
8\pi A_j=\lim_{r\rightarrow\infty}\oint dS_.^i(K_{ij}-\de_{ij}K)\nonumber
\ee
using the fundamental form \ref{eq:A2.5} gives
\be
8\pi A_j=\lim_{r\rightarrow\infty}(B/(B\sqrt{C}),0,0)\nonumber
\ee
and this vanishes in all the cases considered here.
\section{A Field Index}
\label{sec:D}
A perfect fluid has Lagrangian $L_d=p$ (the pressure),  see for example 
Roberts (1996)\cite{bi:mdr96},  and Hamiltonian density $H_d=\mu$ 
(the density).   Taking metric variations of the Lagrangian produces 
the metric stress
\be
T_{ab}=(\mu+p)V_aV_b+p~g_{ab},~~~  V_a V_.^a=-1,
\label{eq:A4.1}
\ee
The absolute derivative is defined by
\be
\dot{X}_{abc\ldots}=V_.^eX_{abc\ldots;e},
\label{eq:A4.2}
\ee
The Bianchi identities give
\ber
&&-V_aT_{..;b}^{ab}=\dot{\mu}+(\mu+p)\ck{\Th},~~~\ck{\Th}=V^a_.,
                                                         \label{eq:A4.3}\\
&&h_{ab}T^{bc}_{..;c}=(\mu+p)_a+h_{a.}^{~b}p_b.\nonumber
\ear
The Eisenhart-Synge fluid index is
\be
\om=ln I=\int\fr{dp}{\mu+p},
\label{eq:A4.4}
\ee
in the literature sometime $I$ is called the index and sometimes $\om$.
Assuming an equation of state 
\be
p=f(\mu),~~~\mu=f^{-1}(p),
\label{eq:A4.5}
\ee
the Bianchi identities can be expressed as
\be
\ck{\Th}=\dot{\om}\fr{df^{-1}}{dp},~~~\dot{V}_a=-h^{~b}_{a.}\om_b.
\label{eq:A4.6}
\ee
For the equation of state \ref{eq:2.5} the index is $\om=[(\ga-1)/\ga]ln\mu$.
An O-scalar field has this equation of state with $\ga=2$ and  
$V_a=\ph_a/sqrt{-\ph_c^2},  p=\mu=-\ph_c^2$.   
Taking a time-like vector field,  say
\be
V_a=(N,0,0,0),~~~   V_.^a=(-1/N,0,0,0),
\label{eq:A4.7}
\ee
it is possible to relate the behaviour of the lapse $N$ to the fluid index 
$\om$.    For many configurations they can be equated $N=g(\om)$,  typically 
$N=1/\om$.    Thus if the fluid index is well-behaved throughout the space it 
is possible,  without recourse to field equations,  to discuss whether the 
metric is.   Other choices of time-like vector-field can be made,  for example
an asymptotically flat space-time there is the normal vector to the 
three-sphere at infinity.

There is no direct analog of the preceeding for fields with infinite degrees 
of freedom.   This is because the stress for fields does not have an explicit 
dependence on vector fields.   It is possible to introduce vector fields into 
an action and vary it to produce an extreme configuration between the metric,
fields,  and vector field,  in the following manner.   Let the fields be 
describable by a Lagrangian
\be
I_l=\int\sqrt{-g}d^4x\Lg(\ph,\ph_a),
\label{eq:A4.8}
\ee
Under infinitesimal coordinate variations this gives
\be
\fr{\de I}{\de x_.^a}:\int\sqrt{-g}d^4x\na\Th_c^{ab},
\label{eq:A4.9}
\ee
where $\Th_c^{ab}$ is the canonical stress
\be
\Th_c^{ab}=\fr{\p\Lg}{\p\ph_a}\p^b\ph-g^{ab}_{..}\Lg_d.
\label{eq:A4.10}
\ee
The Hamiltonian density is usually defined in terms of components 
$\Hd_d=\Th^{~0}_{c.0}$,  more generally it can be defined as
\be
\Hd_d=V_aV_b\Th_c^{ab},
\label{eq:A4.11}
\ee
where $V_a$ is a unit time-like vector-field.   Variations of $I_l$   
with respect to the metric produce the stress
\be
T_{ab}=D_{ab}+\Lg~g_{ab}, 
\label{eq:A4.12}
\ee
where $D_{ab}$ is given by $\de I_d/\de g_{ab}$.   Applying the time-like 
vector-field and using \ref{eq:A4.11} gives
\be
\Hd_d+\Lg_d=V_.^aV_.^bD_{ab}. 
\label{eq:A4.13}
\ee
Variations of actions corresponding to $\Hd_d$ and $\Lg_d$ give dynamical 
equations,   this suggests considering a new action $I_n$ which is a sum of 
the Hamiltonian and Lagrangian actions
\be
I_n=\int\sqrt{-g}d^4x(\Hd_d+\Lg_d)=\int\sqrt{-g}d^4xV_.^aV_.^bD_{ab},
\label{eq:A4.14}
\ee
which will give extremeal (maximally stable or unstable) configurations.   
Another possible way of producing further equations between the fields is to 
consider higher order variations.   For example Bazanski (1977)[23]
\cite{bi:bazanski},  the second covariant of the point particle action 
gives the geodesic deviation equations.

A scalar-electrodynamic Lagrangian for a complex scalar field $\ps$ and a 
vector field $A_a$ is
\ber
&&\Lg=-\fr{1}{4}F^2+\fr{1}{2}\mu^2A^2-\la(A^a_{.a})^2+J_aA_. ^a 
    -{\Dr}_a\ps{\Dr}^a\br{\ps}-V(\ps^2)+2\bt\ps^2R,\nonumber\\
&&{\Dr}_a\ps=\p\ps+ieA_a,~~~\ps^2=\ps\br{\ps},\label{eq:A4.15}\\
&&{\Dr}_a\br{\ps}=\p\br{\ps}-ie\br{\ps}A_a.\nonumber
\ear
Varying with respect to the metric gives
\be
D_{ab}=F_{ac}F^{~c}_{b.}-\mu^2A_aA_b-2J_aA_b+4\la^2A_{ab}A^{~c}_{c.}  
       +2{\Dr}_{(a}\ps\\dr_{b)}\br{\ps}-2\bt(\ps^2)_{ab}-2\bt\ps^2R.
\label{eq:A4.16}
\ee
Setting up the new action $I_n$ and varying with respect to the fields gives
\ber
&&\fr{\de I_n}{\de\ps}:=-2(V^a_.V^b_.{\Dr})b\br{\ps})_a 
                        +2ieA_aV^a_.V^b_.{\Dr}_b\br{\ps}\nonumber\\
&&~~~~~~~~~~           +2\bt\br{\ps}[(V^a_.V^b_.)_{ab}-R_{ab}V^a_.V^b_.],\\
&&\fr{\de I_n}{\de A_a}:=-2(V_aV^b_.F^{~c}_{b.}+V^c_.V^b_.F_{ba})_c  
                      +2V_A(-\mu^2V^b_.A_b-J_bA^b_.)\nonumber\\
&&~~~~~~~~~~            +ie(\ps{\Dr}_a\br{\ps}-\br{\ps}{\Dr}_a\ps),\nonumber
\label{eq:A4.17}
\ear
It is also possible to vary with respect to $\dt{\ps}, \dt{A}_a, g_{ab}$, 
and $V_a$.   Variations with respect to $V_a$ are best done using velocity 
potentials,  see Roberts (1996)[10]\cite{bi:mdr96}.  Varying with respect 
to $g_{ab}$ a new "stress" tensor can be created.  Applying the Bianchi 
identities to it the equations \ref{eq:A4.17} are not recovered,  because the 
conservation equations as derived from \ref{eq:A4.9} are also changed.   
Choosing $D_{ab}=R_{ab}$ and varying with respect to $g_{ab}$ does not 
reproduce Raychaudhuri's equations but instead 
$\Th=R_{ab}V^a_.V^b_.+\Th^2-\dt{V}^a_{;a.}+\fr{1}{2}\Bx V^2$.   For the 
Reissner-Nordstrom solution with vector-field \ref{eq:A4.7},  and the null 
vector field $l_a$,  \ref{eq:A4.17} becomes
\ber
&&\ps_t=-\fr{4\sq{2}}{r^3N^3}e
              \left[1+\fr{2m}{r}-\fr{3e^2}{r}\right],\nonumber\\ 
&&\ps_v=4\sq{2}\fr{e}{r^3}.
\label{eq:A4.18}
\ear
For static O-scalar fields the field index vanishes everywhere,  
for non-static O-scalar fields using $l_a$ gives
\be
\ps=-2\left(\fr{2S'}{S}+\fr{Y'}{Y}\right)\ph'.
\label{eq:A4.19}
\ee
In the solution \ref{eq:4.5} this is
\be
\ps=-\fr{4\si v}{Y^2}(r-\si v).
\label{eq:A4.20}
\ee

\end{document}